\documentclass[a4paper]{article}

\usepackage{INTERSPEECH2020}
\usepackage{subfigure}
\usepackage{multirow}
\usepackage{array}
\usepackage{color}
\newcolumntype{P}[1]{>{\centering\arraybackslash}p{#1}}
\usepackage{array}
\usepackage{booktabs}

\title{Noise Tokens: Learning Neural Noise Templates for Environment-Aware Speech Enhancement}

\name{Haoyu Li$^1{^,}^2$, Junichi Yamagishi$^1{^,}^2$}

\address{ 
  $^1$National Institute of Informatics, Japan \quad $^2$SOKENDAI, Japan}
\email{haoyuli@nii.ac.jp, jyamagis@nii.ac.jp}

\begin{document}

\maketitle
\begin{abstract}
  In recent years, speech enhancement (SE) has achieved impressive progress with the success of deep neural networks (DNNs). However, the DNN approach usually fails to generalize well to unseen environmental noise that is not included in the training.
  To address this problem, we propose ``noise tokens" (NTs), which are a set of neural noise templates that are jointly trained with the SE system. NTs dynamically capture the environment variability and thus enable the DNN model to handle various environments to produce STFT magnitude with higher quality. 
  Experimental results show that using NTs is an effective strategy that consistently improves the generalization ability of SE systems across different DNN architectures. Furthermore, we investigate applying a state-of-the-art neural vocoder to generate waveform instead of traditional inverse STFT (ISTFT). Subjective listening tests show the residual noise can be significantly suppressed through mel-spectrogram correction and vocoder-based waveform synthesis.

\end{abstract}

\section{Introduction}

Speech enhancement (SE) aims to improve the intelligibility and quality of degraded speech by suppressing noise. It has been widely used as preprocessors in speech-related applications, such as hearing aids, speech communication, and automatic speech recognition (ASR).
Recently, deep neural network (DNN) based SE has become the mainstream approach and made notable progresses \cite{xu2014regression, weninger2015speech}. However, the DNN approach is inevitably limited by its generalization ability to unseen noises. Real-world environment variability is much more complex, and it is hard for the DNN to model it sufficiently well. The mismatch between training and real-world environments leads to serious performance degradation.

To address this problem, we introduce ``noise tokens'' (NTs) into the SE system. The noise tokens (NTs) model is inspired by recent progress \cite{wang2018style} in expressive speech synthesis, where conceptually-similar ``style tokens'' were proposed to model acoustic expressiveness to control speaking style. We adapt and revise the original style tokens approach to the SE task. In particular, a noise token layer (NTL) combined with a set of trainable noise templates is designed to dynamically extract noise embedding from the input noisy speech. The NTL projects the noisy speech onto a \textit{noise subspace} by assigning weights to each noise template. Noise embedding is obtained with the weighted sum of these templates and then jointly trained with the SE system. By factorizing unseen noises into a linear combination of learned templates, we expect that the DNN model can handle various environments in a more efficient way. In addition, to alleviate the phase distortion and further suppress the residual noise, we use a mel-spectrogram prediction network (MPN) to predict the clean mel-spectrogram from the enhanced STFT magnitude. \mbox{WaveRNN} \cite{kalchbrenner2018efficient} vocoder is then applied to generate the final waveform with significantly better noise reduction compared to the ISTFT-based counterpart. 
\vspace{-2mm}
\subsection{Related work}
\label{sec:related-work}
In \cite{xu2014dynamic}, a dynamic noise aware training (DNAT) technique was proposed to enrich the DNN's generalization, where the noise embedding is performed in advance by using either a traditional noise tracking algorithm or IBM-based estimation \cite{wang2014training}. In this context, however, the noise embedding generated with a separate noise estimation module might be suboptimal and inefficient. In our work, the noise embedding produced by NTs is jointly optimized with the enhancement DNN model, which facilitates flexibility and the effectiveness of the whole system. 

The authors in \cite{maiti2019speaker} firstly proposed applying neural vocoders for parametric resynthesis SE. Compared to their work, we do not focus on investigating the effect of different neural vocoders for the synthesis module but only focus on the enhancement module. In our work, an MPN with an auto-regressive mechanism is designed as a post-processor to predict the clean mel-spectrogram from the enhanced STFT magnitude. In contrast, only a simple prediction model was adopted in \cite{maiti2019speaker} to predict clean acoustic features from noisy speech directly. As for waveform synthesis, we simply selected WaveRNN as vocoder.
\vspace{-1.86mm}
\section{System Overview}
\vspace{-0.2mm}
\begin{figure}[tbp]
  \centering
    \includegraphics[height=142.5pt,width=1\linewidth]{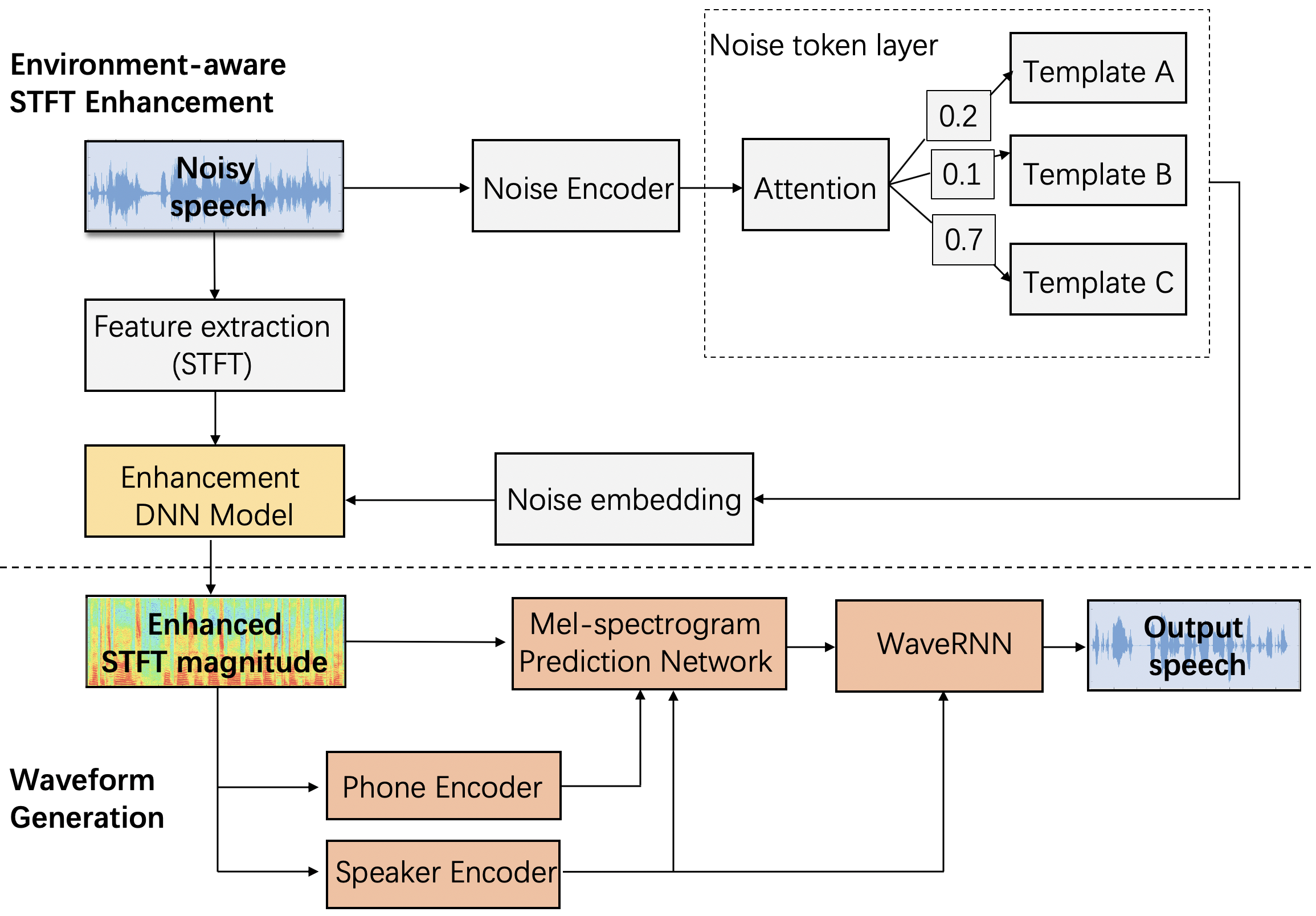}
  \caption{System diagram of proposed SE model.}
  \label{fig:diagram}
  \vspace{-6mm}
\end{figure}

The system diagram is illustrated in Figure~\ref{fig:diagram}. Our proposed system consists of two parts. First, an enhancement model produces the enhanced STFT magnitude from the noisy speech. Second, we generate the waveform from the STFT magnitude by using WaveRNN vocoder.

\subsection{Environment-aware STFT enhancement}
\vspace{-1mm}
The first part estimates the enhanced magnitude by the enhancement model together with the proposed noise tokens (NTs). 
\vspace{-2mm}

\subsubsection{Noise tokens model}
Compared to conventional DNN-based pipeline, NTs are introduced in this part to extract the noise embedding and inform the DNN model of environment information. The NTs model consists of a noise encoder and a noise token layer (NTL).

The noise encoder takes as input the noisy magnitude spectrogram. It is composed of 6 layers of 2-D CNN each with 3 (along the time axis)$\times$3 (along the frequency axis) kernel, 1$\times$2 stride, batch normalization, and ReLU activation. The output channels are set to 32, 32, 64, 64, 128, and 128, respectively.
A bi-directional GRU with 128 nodes is followed by the last CNN layer, resulting in a 256-dimensional (128$\times$2) feature for each time step. The output of the GRU is regarded as an encoded environment representation, which is then passed to the NTL.
The NTL is composed of 16 trainable noise templates (tokens) and a multi-head attention module \cite{vaswani2017attention}. Each template has 256 dimensions, and the number of attention heads is set to 8.
The representation produced by the previous noise encoder is served as the \textit{query} vector here, and the attention module calculates the similarity (weight) between the encoded representation and each template. The noise embedding is then generated as the weighted sum of the noise templates and fed as an additional input into the enhancement model. 

\begin{figure}[tbp]
  \centering
    \includegraphics[width=\linewidth]{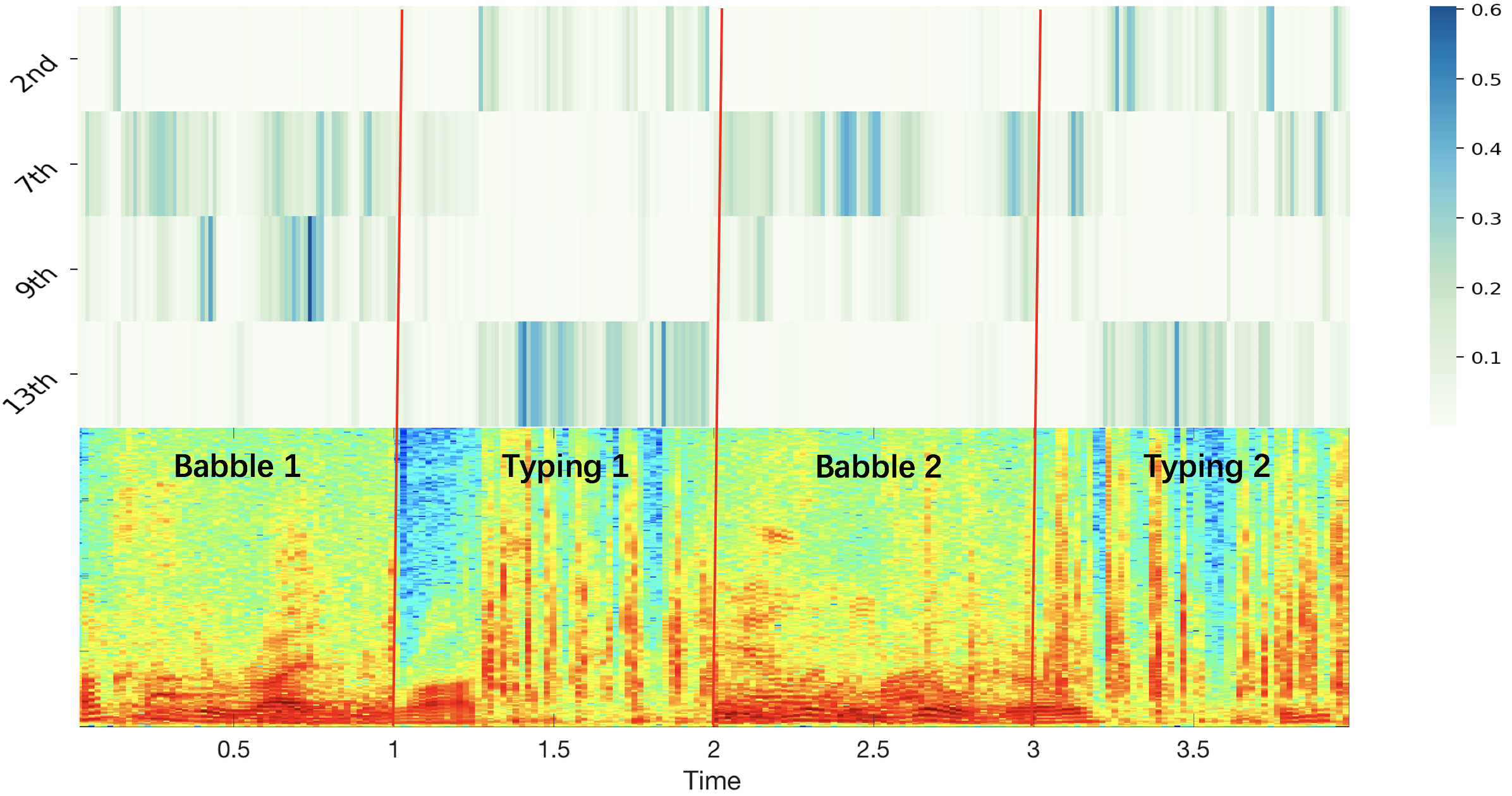}
  \caption{Visualization of template weights for noisy speech where \textit{Babble} and \textit{Typing} noises alternatively appear. There are 16 noise templates for each of the 8 attention heads. For clear visualization, we only list four templates (2\textsuperscript{nd}, 7\textsuperscript{th}, 9\textsuperscript{th}, and 13\textsuperscript{th}) on the first attention head branch. This model is trained with BLSTM architecture on a 50-hour noisy speech data set. The detailed configurations will be discussed in Section~\ref{sec:exp}.}
  \label{fig:visual}
  \vspace{-6mm}
\end{figure}

Unlike \cite{wang2018style} where only a global embedding was considered, we generate noise embedding in a dynamic frame-by-frame manner to fully capture the non-stationary environment information. In addition, since noise templates are jointly trained with the whole system in an unsupervised way, we expect that the learned templates can be representative enough to model various environmental noises. Figure~\ref{fig:visual} gives a visualization example of learned template weights for noisy speech.
We can clearly see that the 7\textsuperscript{th} and 9\textsuperscript{th} noise templates are activated during \textit{Babble} segments, while the 2\textsuperscript{nd} and 13\textsuperscript{th} templates are active during \textit{Typing} segments. This shows that the proposed noise tokens can effectively capture and adapt to varying environments. 

\vspace{-4.5mm}
\subsubsection{Masking-based enhancement model}
\label{sec:semodel}
\vspace{-0.5mm}
We design an enhancement DNN model to enhance the STFT magnitude. The model predicts a real-value soft mask $M$, which is then element-wise multiplied with the noisy spectrogram $S$ to obtain the enhanced spectrogram $\Tilde{S} = M \odot S$. As suggested in \cite{wilson2018exploring}, we adopt the following mean squared error (MSE) as the loss function in the training:
\vspace{-2mm}
\begin{equation}
  L = \left\| |\Tilde{S}^{0.3}|-|S^{0.3}| \right\|^2 + \lambda \left\| \Tilde{S}^{0.3}-S^{0.3} \right\|^2
  \label{eq:1}
\end{equation}
where the MSE of both magnitude and complex spectra are taken into account, with a weight parameter $\lambda=0.1$. Although our target is the enhanced magnitude, i.e., $|\Tilde{S}|$, the MSE of the complex spectrogram is integrated into the loss function with the aim of somewhat reducing the phase distortion. All spectral in Equation~(\ref{eq:1}) are power-law compressed with a power of $0.3$. 

We test three different DNN architectures for the enhancement model. The detailed descriptions will be given in Section~\ref{sec:dnnac}.

\vspace{-1.5mm}
\subsection{Waveform generation module}
\vspace{-0.5mm}
The second part generates waveform by using mel-spectrogram prediction network (MPN) and \mbox{WaveRNN} vocoder.

The MPN aims to further suppress the residual noise by predicting the clean mel-spectrogram. Similar to \cite{lorenzo2018can}, it is designed as an auto-regressive model.
The input features are first processed by a feed-forwarding layer with 768 nodes and a BLSTM layer with 400 nodes. A unidirectional LSTM with 256 nodes then takes as input the mel-spectrogram generated in the previous frame and its previous state. Four feed-forwarding layers are successively added, each with 80 nodes, to produce the 80-dimensional mel-spectrogram of the current time step. All feed-forwarding layers use LeakyReLU activation with \mbox{slope $=0.3$}, except for the output layer, which uses a linear function.
For better performance, we further use phone and speaker embeddings (both extracted from the enhanced magnitude) in the MPN to emphasize the content information and speaker characteristics of the processed utterance. 
In particular, the 64-dimensional speaker embedding comes from a pre-trained learnable dictionary encoding (LDE) \cite{cooper2019zero} based speaker verification system. And a phone embedding with 256 dimensions is obtained from the last bottleneck layer of a phone recognition model trained with connectionist temporal classification (CTC) loss \cite{graves2014towards}. We use the open-sourced implementations for the above speaker and phone encoders\footnote[1]{\texttt{https://github.com/Diamondfan/CTC\_pytorch}}\footnote[2]{\texttt{https://github.com/jefflai108/pytorch-kaldi-\\neural-speaker-embeddings}}.
These two auxiliary embeddings are concatenated with the enhanced STFT magnitude as the input features and help the MPN produce the mel-spectrogram with higher quality.

Instead of ISTFT, \mbox{WaveRNN} vocoder tries to directly synthesize high quality waveform by avoiding introducing the noisy phase. In addition to the mel-spectrogram, the extracted speaker embedding is also fed into the vocoder as local conditions for higher synthesis quality. We use the open-sourced \mbox{WaveRNN} implementation\footnote[3]{\texttt{https://github.com/mkotha/WaveRNN}}.
\vspace{-1.5mm}
\section{Experiments}
\label{sec:exp}
\vspace{-1mm}
\subsection{Data preparation}
\vspace{-0.3mm}
The MS-SNSD dataset \cite{reddy2019scalable} is used in our experiments. We select 7 and 4 noise types to prepare the training and test sets, respectively. For the training set, we further add another 14 noise types from Nonspeech sounds database \cite{hu2004100} to expand the diversity in noises. For the test set, the 4 selected noise types are: babble, typing, squeaky chair, airport announcements. As we only study the DNN's generalization to unseen noises, none of these 4 types are included in the training. Finally, a 50-hour training set (around 36,000 audio clips) is generated with 21 noise types at 5 SNR levels, i.e., {-5, 0, 5, 10, 15} dB. The test set consists of 2 hours of noisy speech, with 4 unseen noise types at 5 SNR levels, i.e., {-2.5, 2.5, 7.5, 12.5, 17.5} dB. In addition, we use the VCTK corpus \cite{yamagishi2019cstr} and TIMIT database \cite{garofolo1993timit} to train \mbox{WaveRNN} vocoder and phone encoder, respectively. All audios used in our experiments are resampled at 16 kHz.
\vspace{-2mm}
\subsection{Performance analysis with noise tokens}
\label{sec:dnnac}
\vspace{-0.5mm}
\begin{table}[t]
    \caption{Average PESQ and STOI scores with different noise embeddings across three DNN architectures under test unseen noises.}
    \label{tab:nts}
    \centering
    \renewcommand\arraystretch{1.4}
    \scalebox{0.808}{
    \begin{tabular}{||c||c|c||c|c||c|c||}
        \hline
          \multirow{2}{*}{Architectures} &
           \multicolumn{2}{c||}{w/o embedding \quad} & \multicolumn{2}{c||}{with DNAT \quad} & \multicolumn{2}{c||}{with NTs \quad}  \\
           \cline{2-7}
           & PESQ & STOI & PESQ & STOI &  PESQ & STOI \\
          \hline
          \hline
          BLSTM & 2.686 & 0.896 & 2.692 & 0.898 & \textbf{2.858} & \textbf{0.914} \\
        VoiceFilter & 2.792 & 0.904 & 2.771 & 0.902 & \textbf{2.907} & \textbf{0.916} \\
        T-GSA & 2.754 & 0.906 & 2.726 & 0.902 & \textbf{2.808} & \textbf{0.912} \\
        \hline
    \end{tabular}
    }
    \vspace{-5mm}

\end{table}
First, we examine if the performance can be improved by using noise tokens (NTs). We systematically test the effectiveness of NTs with three state-of-the-art enhancement DNN architectures: (1) BLSTM, a standard model with 2 BLSTM layers and 1 fully-connected layer; (2) VoiceFilter \cite{wang2018voicefilter}, a CNN-BLSTM model that consists of 8 layers of 2-D CNNs, followed by 1 BLSTM layer and 2 feed-forwarding layers; and (3) T-GSA \cite{kim2019transformer}, a Transformer-based model that has 4 Transformer blocks \cite{vaswani2017attention} with Gaussian-weighted self-attention.

Each architecture is used as the enhancement model and trained with or without the noise embedding. Moreover, we also compare our proposed NTs approach with the dynamic noise aware training (DNAT) method, where the noise power spectral density (PSD) estimated by a noise tracking algorithm \cite{rangachari2006noise} was regarded as the noise embedding.
Since we only focus on the performance of NTs, but not on the waveform generation (WG) module in this preliminary experiment, we apply ISTFT to generate the waveform with the noisy phase instead of using the WG module. The PESQ \cite{rix2001perceptual} and STOI \cite{taal2011algorithm} scores are used as objective measures. The experimental results presented in Table~\ref{tab:nts} show that using the proposed NTs is a universal and effective technique that consistently improves the generalization ability of SE systems across all three tested architectures. Using NTs also outperforms the DNAT method, which demonstrates that the neural noise embedding is more efficient than the signal-processing based noise estimations. 
\vspace{-2mm}
\subsection{Impact of noise diversity}
\vspace{-0.5mm}
The generalization of SE systems can be improved by feeding them with a diverse noise corpus with more noise types. In this experiment, we analyze the impact of noise diversity on performance. The original training noise corpus (with 21 noise types) is divided into three smaller subsets, each with \{7, 12, 16\} noise types. Thus, we have 4 noise corpora (N7, N12, N16, N21) in total, and each is used to generate a 50-hour training set. Note that these four generated training sets share the same configurations, i.e., the clean speech data set, size of noisy speech data (all are 50 hours in duration), and SNR levels, while they only differ from each other in the number of noise types they are mixed with. We use the standard BLSTM architecture described in Section~\ref{sec:dnnac} as the enhancement model and apply ISTFT for waveform synthesis. Systems with and without NTs are trained with 4 noise corpora, respectively, and then tested on the same test set. The PESQ results are given in Table~\ref{tab:diversity}. We can see that feeding more noise types into training always helps improve the performances of both systems. Compared to the system without noise embedding, NTs bring higher relative improvements on PESQ with increasing noise diversity, which indicates that the proposed NTs can effectively exploit multiple noises due to the modelling ability of their trainable templates.
\begin{table}[t]
    \caption{Average PESQ score and its relative improvements with different training noise corpora under test unseen noises.}
    \label{tab:diversity}
    \centering
    \renewcommand\arraystretch{1.3}
    \scalebox{0.81}{
    \begin{tabular}{||c||p{30pt}<{\centering}|c||p{30pt}<{\centering}|c||}
        \hline
          \multirow{2}{*}{Noise corpus} &
           \multicolumn{2}{c||}{BLSTM w/o NTs \quad} & \multicolumn{2}{c||}{BLSTM with NTs \quad}  \\
           \cline{2-5}
           & PESQ & Relative imp. & PESQ & Relative imp. \\
          \hline
          \hline
          N7 & 2.564 & 0.00\% & \textbf{2.657} &\textbf{ 0.00\%} \\
        N12 & 2.639 & 2.94\% & \textbf{2.786} & \textbf{4.86\%} \\
        N16 & 2.672 & 4.20\% & \textbf{2.812} & \textbf{5.82\%} \\
        N21 & 2.685 & 4.71\% & \textbf{2.858} & \textbf{7.54\%} \\
        \hline
    \end{tabular}
    }
\end{table}
\vspace{-2mm}
\subsection{Waveform generation module}
\label{sec:wgm}
\vspace{-0.5mm}
Next, we check to determine if the waveform generation module shown in Figure~\ref{fig:diagram} can synthesize higher quality speech with better noise reduction. The enhanced STFT magnitude is first obtained from the enhancement DNN model implemented with the BLSTM architecture with NTs. The enhanced STFT magnitude can then be converted to the waveform by either ISTFT or the waveform generation (WG) module. We denote the systems with these two methods as \texttt{NTs-ISTFT} and \texttt{NTs-WG}. Noisy speech without any processing is also compared. Table~\ref{tab:wgobj} gives the objective results and shows that the proposed WG module is much worse than the traditional ISTFT. The probable reason is that the PESQ and STOI are not designed to evaluate neural vocoders, which also explains why these measures are not typically used in the field of speech synthesis. Such results further encourage us to conduct the following subjective listening tests.
\begin{table}[t]
    \caption{Average PESQ and STOI scores with different waveform synthesis methods under test unseen noises.}
    \label{tab:wgobj}
    \centering
    \renewcommand\arraystretch{1.3}
    \scalebox{0.85}{
    \begin{tabular}{||p{60pt}<{\centering}||p{50pt}<{\centering}|p{50pt}<{\centering}||}
        \hline
          Methods & PESQ & STOI \\
          \hline
          \hline
          Noisy & 2.021 & 0.833 \\
          \hline
          \hline
        NTs-ISTFT & \textbf{2.858} & \textbf{0.914}  \\
        NTs-WG & 2.509 & 0.867  \\
        \hline
    \end{tabular}
    }
    \vspace{-4.3mm}
\end{table}
\vspace{-5mm}
\subsection{Listening tests}
\vspace{-0.5mm}
Next, crowdsourced listening tests are conducted to comprehensively evaluate different systems. Since the WG module can be directly applied to the noisy speech to generate waveform, it is also included as a tested system. We summarize the notations for the systems we evaluated in the listening tests:
\begin{itemize}
    \item \textbf{Baseline}: BLSTM model without noise tokens. ISTFT is used to generate waveform.
    \item \textbf{NTs-ISTFT}: BLSTM model with noise tokens. ISTFT is used to generate waveform.
    \item \textbf{NTs-WG}: BLSTM model with noise tokens. Waveform generation module is then used to generate waveform from the enhanced STFT magnitude.
    \item \textbf{WG}: Waveform generation module is directly applied to the input noisy STFT magnitude to generate waveform.
    \item \textbf{Clean}: Raw clean speech without noise.
    \item \textbf{Noisy}: Degraded noisy speech without any processing.
\end{itemize}
We choose 96 files from the test set for each system\footnote[4]{Audio samples of the tested files are available at: \texttt{https://\\nii-yamagishilab.github.io/samples-NTs/}}. Subjects were asked to rate the speech quality, noise suppression, and the overall performance of the anonymized file from 1-5 for the mean opinion score (MOS). 
For reference, the clean and noisy versions of each file were also provided to subjects before rating. 
Each file was rated ten times in order to avoid human bias, and 521 subjects participated. To reduce the burden on the subjects, the test files that were more than 12 seconds in duration were manually split into smaller segments of at most 5 seconds. 

\begin{figure}[t]
  \centering
  \subfigure[Results on speech quality]{
    \includegraphics[width=0.98\linewidth,height=88.5pt]{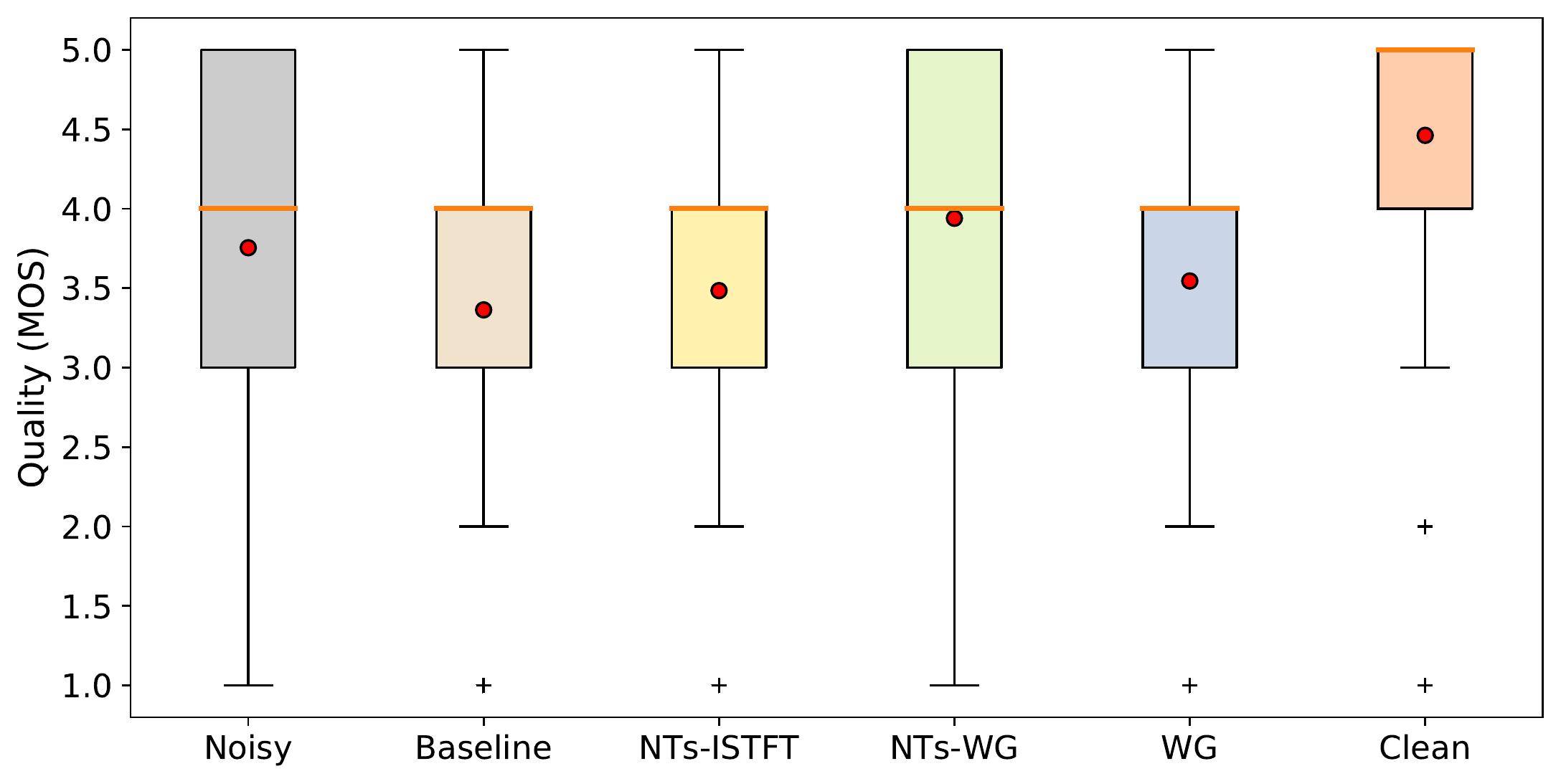}
  }    
  \vspace{-3mm}

  \subfigure[Results on noise suppression]{
    \includegraphics[width=0.98\linewidth,height=88.5pt]{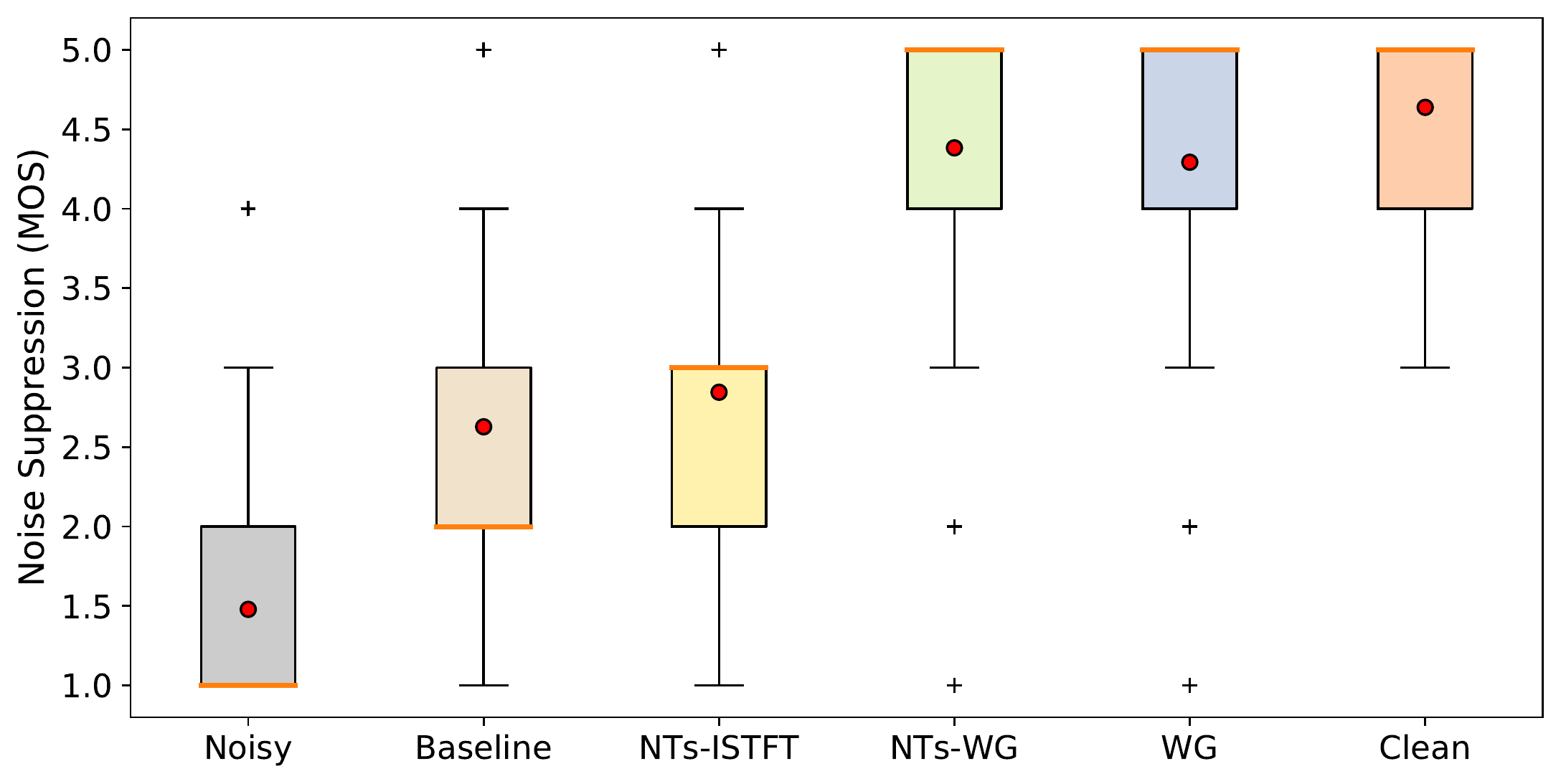}
  }
      
      \vspace{-3mm}

  \subfigure[Results on overall performance]{
    \includegraphics[width=0.98\linewidth,height=88.5pt]{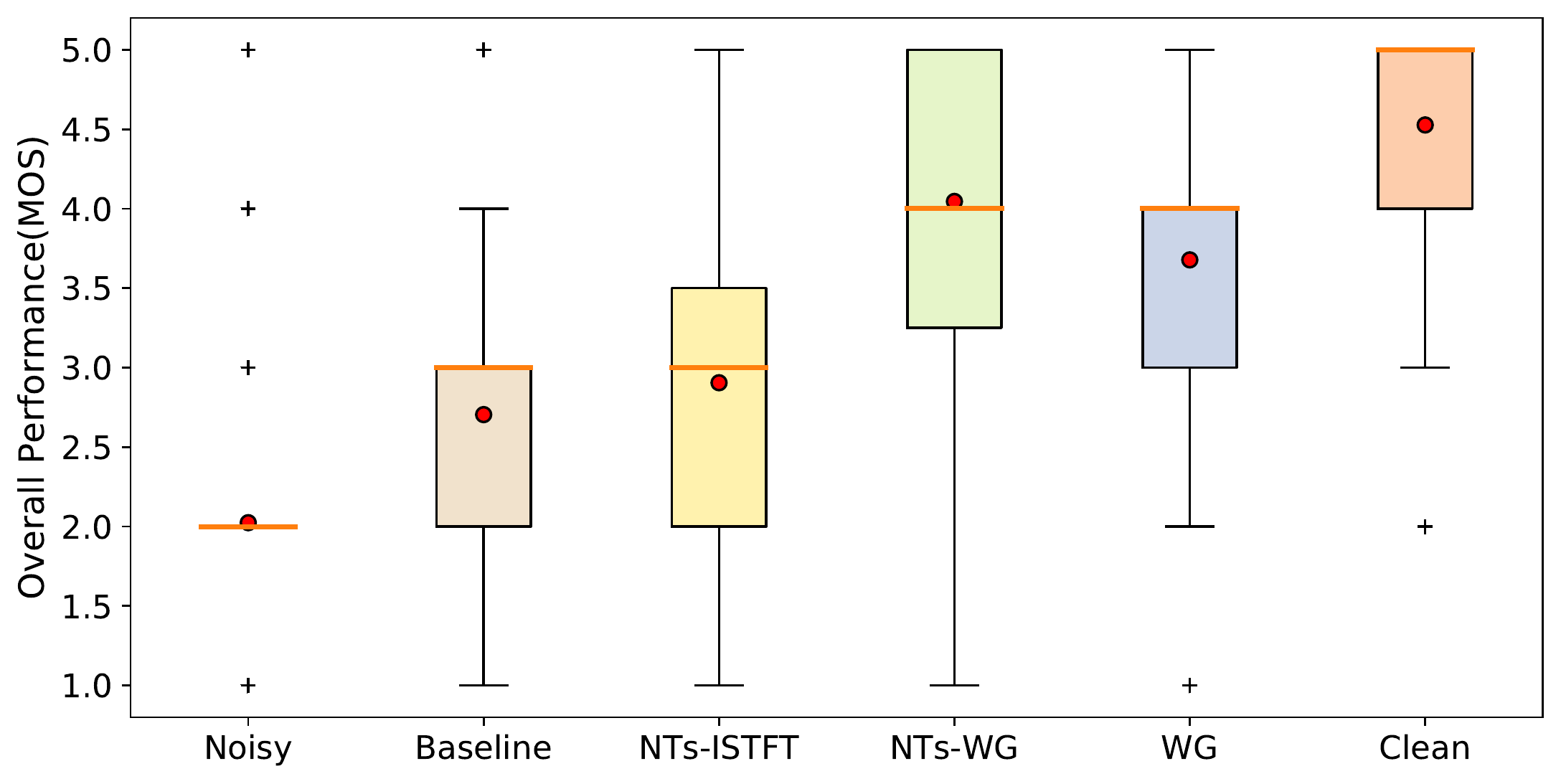}
  }
    \vspace{-4mm}
  \caption{Box plots on speech quality, noise suppression and overall performance. Red dots represent mean score.}
  \label{fig:sbjres}
  \vspace{-6mm}
\end{figure}

\begin{figure}[t]
  \centering
  
  \subfigure[Clean]{
    \includegraphics[width=0.45\linewidth,height=77.7pt]{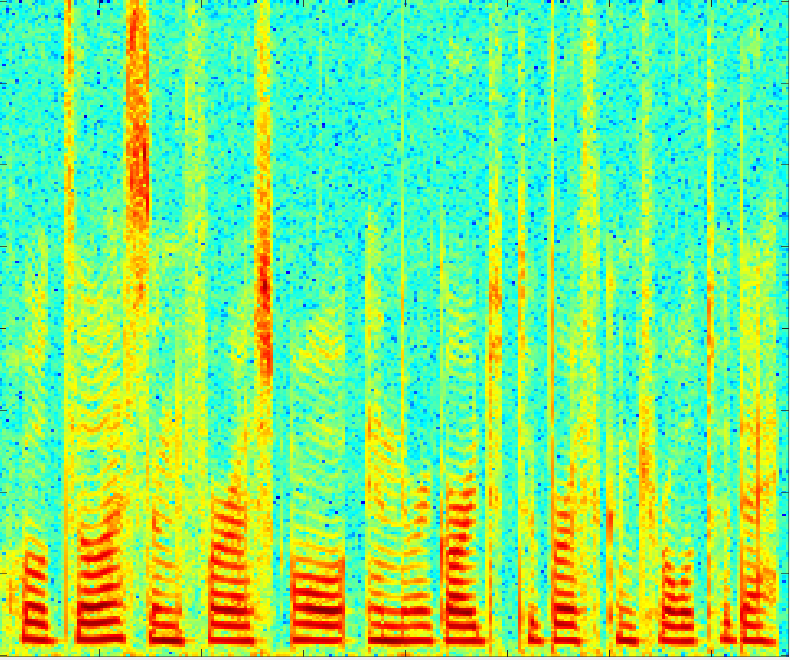}
  }
  \subfigure[Noisy]{
    \includegraphics[width=0.45\linewidth,height=77.7pt]{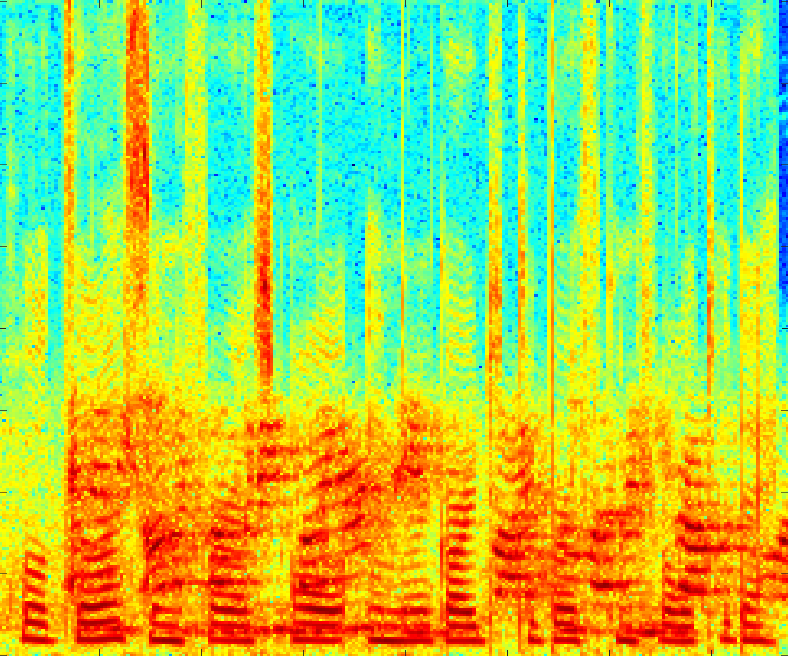}
  }
  \vspace{-3mm}
    \subfigure[Baseline]{
    \includegraphics[width=0.45\linewidth,height=77.7pt]{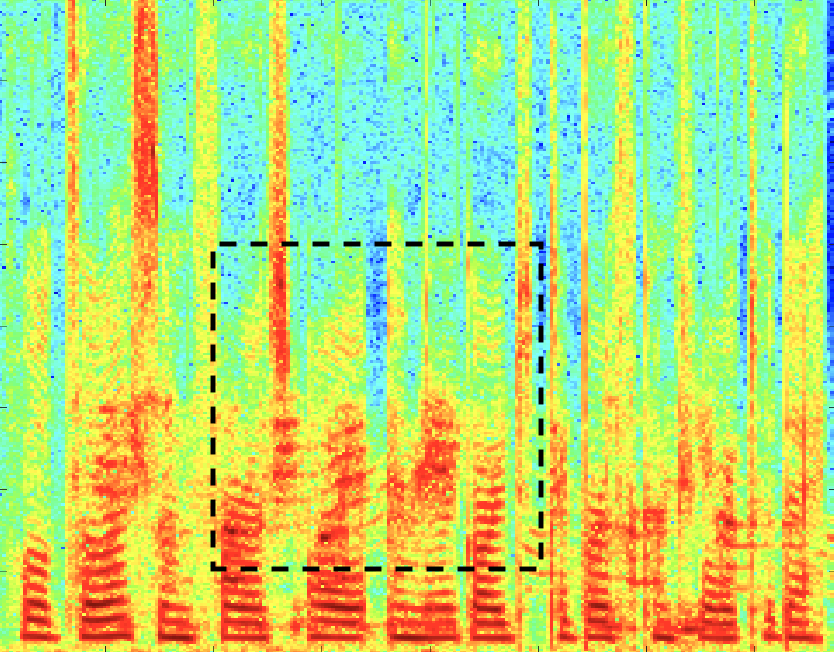}
  }
    \subfigure[NTs-ISTFT]{
    \includegraphics[width=0.45\linewidth,height=77.7pt]{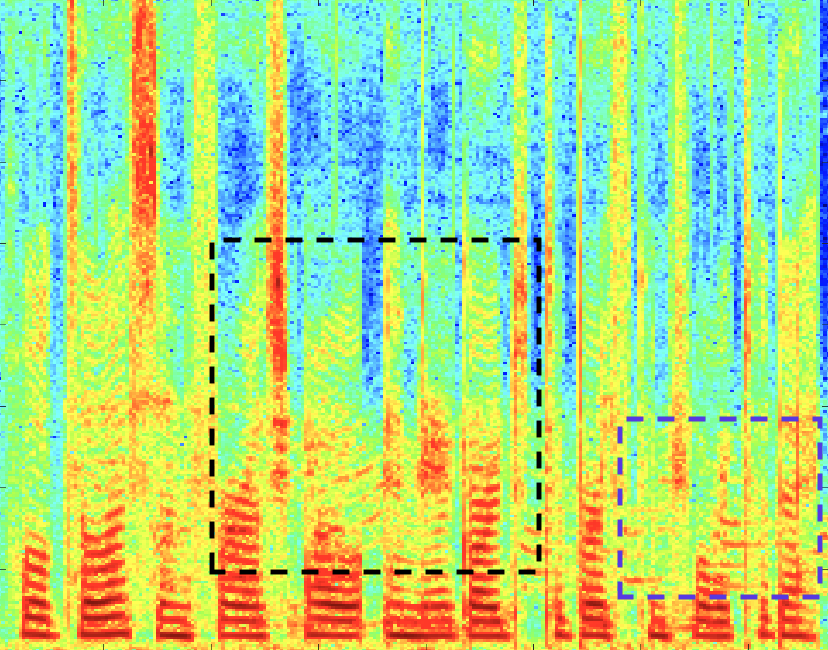}
  }
    \vspace{-3mm}
    \subfigure[NTs-WG]{
    \includegraphics[width=0.45\linewidth,height=77.7pt]{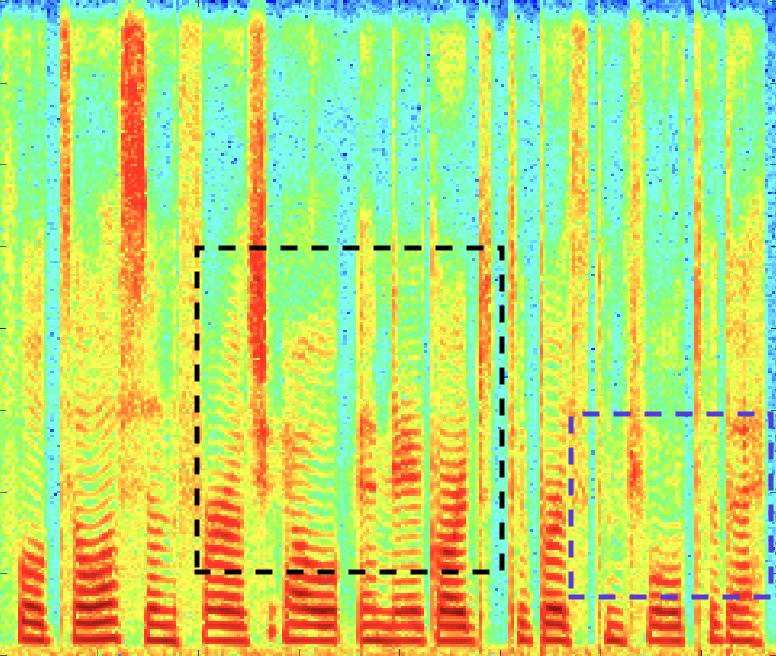}
  }
    \subfigure[WG]{
    \includegraphics[width=0.45\linewidth,height=77.7pt]{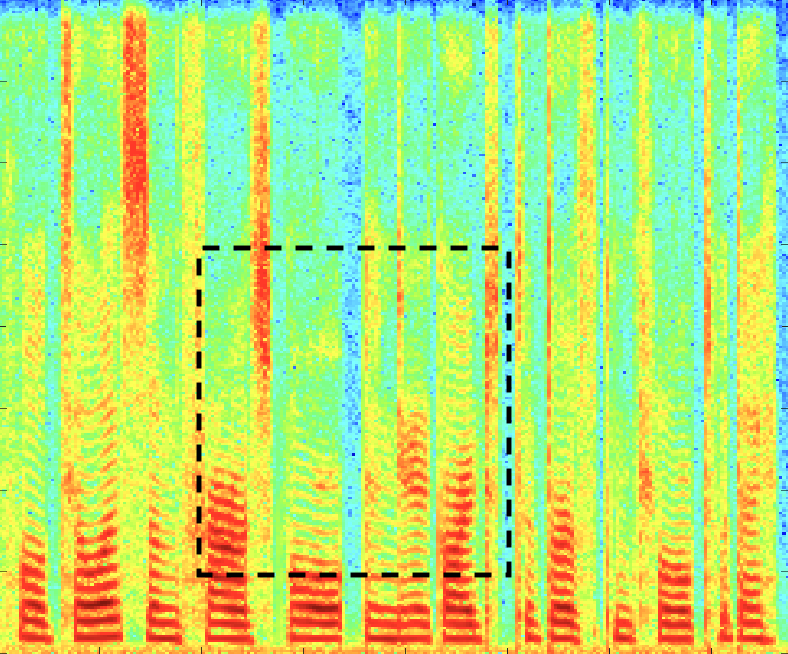}
    \vspace{-3mm}
  }
  \caption{Examples of spectrograms under airport announcement noise at 2.5dB for different systems: (a) Clean, (b) Noisy, (c) Baseline, (d) NTs-ISTFT, (e) NTs-WG, (f) WG.}
  \label{fig:examples}
  \vspace{-6mm}
\end{figure}

The subjective results are shown in Figure~\ref{fig:sbjres}. The Mann-Whitney U test \cite{nachar2008mann} reveals that the \texttt{NTs-ISTFT} system outperforms \texttt{Baseline} in all three scores with \textit{p}-values all lower than 0.005, which demonstrates the effectiveness of noise tokens. 
Compared to \texttt{NTs-ISTFT}, \texttt{NTs-WG} shows significantly higher performances, especially on the noise suppression score. This indicates that the waveform generation module successfully improves the quality and further suppresses the residual noise. 
Furthermore, \texttt{NTs-WG} outperforms \texttt{WG}, which means our proposed two-round enhancement framework, where the WG module is used as a post-processor, is better than the method that applies only the WG module to the noisy input. 
Finally, examples of enhanced spectrograms of the evaluated systems are given in Figure~\ref{fig:examples}. We can clearly see that noises are more suppressed for the \texttt{NTs-ISTFT} system than for \texttt{Baseline}. Compared to the clean reference, some acoustic artifacts in the middle frequency part are introduced by the vocoder-based \texttt{NTs-WG} and \texttt{WG} systems. However, these artifacts do not affect human perception, while the residual noises in \texttt{NTs-ISTFT} are further removed after using a post-processing waveform generation module.

\vspace{-2mm}
\subsection{Discussions}
\vspace{-0.5mm}
 Although \texttt{NTs-WG} shows bad results in terms of PESQ and STOI, it performs best in the subjective listening tests. This indicates that the acoustic artifacts introduced by neural vocoders degrade the objective performances but do not affect the human subjective evaluations. However, we still find that the performances of the WG module are not perfectly stable. This problem occurs with a limited number of cases, but some vocoder-generated samples (from \texttt{NTs-WG} and \texttt{WG} systems) are seriously distorted and thus have very bad quality. This also explains the unsatisfactory lower whisker of the \texttt{NTs-WG} system. 
 To improve the robustness of the vocoder-generated speech is our future work. In addition, we find that the raw \texttt{WG} system can even outperform \texttt{NTs-ISTFT}, which means the waveform generation module itself can be seen as a powerful enhancement model. 
 We plan to further study the WG module and integrate it with the proposed noise tokens.
  \vspace{-2mm}

\section{Conclusion}
\vspace{-0.58mm}
In this paper, we propose noise token model to alleviate the noise mismatch problem of DNN-based SE systems. The neural noise embedding, that is made up of trainable noise templates, can dynamically capture the environment information and thus enriches the DNN's generalization. Experimental results show that the noise token model is effective across various DNN architectures and has higher performance growth with increasing noise diversity. Moreover, we further apply WaveRNN vocoder to synthesize the waveform instead of traditional ISTFT. Subjective listening tests show that the residual noise can be significantly reduced by the proposed waveform generation module.

\vspace{0.1mm}
\noindent
\textbf{Acknowledgments}
This work was partially supported by a JST CREST Grant (JPMJCR18A6, VoicePersonae project), Japan, and by MEXT KAKENHI Grants (16H06302, 17H04687, 18H04120, 18H04112, 18KT0051, 19K24372), Japan. The numerical calculations were carried out on the TSUBAME 3.0 supercomputer at the Tokyo Institute of Technology.

\bibliography{template.bbl}

\end{document}